# An exact stochastic mean-field approach to the fermionic many-body problem

O. Juillet [1], Ph. Chomaz[2], D. Lacroix[1], F. Gulminelli[1]

[1] *LPC/ISMRA, Boulevard du Marechal Juin, F-14050 Caen Cedex , France*

[2] *Grand Accélérateur National d'Ions Lourds, BP 5027, F-14076 Caen Cedex 5, France*

**Abstract** :

We investigate a reformulation of the dynamics of interacting fermion systems in terms of a stochastic extension of Time Dependent Hartree-Fock equations. The noise is found from a path-integral representation of the evolution operator and allows to interpret the exact N-body state as a coherent average over Slater determinants evolving under the random mean-fied. The full density operator and the expectation value of any observable are then reconstructed using pairs of stochastic uncorrelated wave functions. The imaginary time propagation is also presented and gives a similar stochastic one-body scheme which converges to the exact ground state without developing a sign problem. In addition, the growth of statistical errors is examined to show that the stochastic formulation never explode in a finite dimensional one-body space. Finally, we consider initially correlated systems and present some numerical implementations in exactly soluble models to analyse the precision and the stability of the approach in practical cases.

# I - Introduction

Even with present-day computing facilities, the theoretical study of the structure and the dynamics of many-fermion systems, like nuclei, atomic clusters or quantum dots, remains a formidable task which requires approximations to step down to feasibility. In such a context, Hartree-Fock (HF) theory [1] is usually considered as the basic tool. The complex many-body problem is then reduced to an effective single-particle description in which the interaction is communicated through a common and self consistent mean-field. In nuclear physics, static HF calculations provide a very good starting point to investigate many properties of the ground state [2]. The time-dependent model (TDHF) is also successful at low energy in heavy-ion reactions [2] and for the nonlinear electron dynamics in metal clusters [3].

Nevertheless, the HF description only treats in average the two-body potential and thus needs to be improved for strongly correlated systems where the residual interaction can generate coherent superpositions of many mean-fields and gives rise to collision-type processes in the dynamical evolution. In consequence, various schemes have been developed to take into account correlation effects beyond the HF approximation. For example, in the condensed matter problem and in nuclear spectroscopy, recent developments make use of a path integral representation of the Boltzmann operator to select, in a stochastic way, mean-fields dominating the structure of low-lying states [4,5]. Concerning dynamical problems, extensions of TDHF are also possible by adding a Boltzmann-like collision term deduced from an appropriate truncation in the hierarchy of equations describing the time-evolution of reduced density-matrix [6]. However, realistic applications of such a quantum transport theory remain difficult due to the large number of rarely occupied orbitals needed in the simulation [7]. Most of collisional mean-field calculations have thus been carried out in the semi-classical limit via an integro-differential equation for the one-body phase space distribution, known as the Boltzmann (or Vlasov)-Ühling-Uhlenbeck (BUU or VUU) model [8]. In nuclear physics as in the context of metal clusters, many interesting studies of medium-energy processes have been done with such a transport equation concerning, for example, thermalization induced by binary collisions [9] or energy tranfers [10]. Nevertheless, the introduction of the Pauli blocked collision term becomes insufficient when fluctuations are crucial such as in the formation of intermediate mass fragments in heavy-ion reactions [11]. To investigate these situations, the BUU scheme can be improved with the help of a stochastic force, leading to the Boltzmann-Langevin equation (BLE) [12] which has given very promising results in the study of a possible connection between multifragmentation phenomena and the spinodal decomposition of nuclear matter [13]. In addition, the BLE model can be derived in the context of various quantal frameworks [14] and in particular from a stochastic time-dependent Hartree Fock theory [15]. After a sufficiently large time interval, this approach assumes that the full density can be reduced to a statistical mixing of Slater determinants moving in their own mean-field, each of them having a probability given by the residual interaction via Fermi's Golden rule. Such a scheme has only been applied in a BUU-type evaluation of the jumps between TDHF trajectories [16] and its validity remains to be shown in practical cases.

The aim of this Letter is to present a new formulation of the many-fermion problem in terms of mean-field equations with a 1 particle - 1 hole (1p-1h) gaussian white noise. The stochastic one-body process thus constructed will have an average identical to the exact solution and can be used for spectroscopic or dynamical studies as illustrated in exactly soluble models. Statistical errors are moreover analytically calculated to show the stability of the proposed scheme.

## II- From the many-body problem to an exact stochastic mean-field formulation

To begin, let us consider the building of correlations on a Slater state $|\Psi\rangle$ during a small time-step $\Delta t$ and under a general Hamiltonian $H = \sum_{ij} T_{ij}\, a_i^+ a_j - \frac{1}{4}\sum_{ijkl} V_{ijkl}\, a_i^+ a_j^+ a_k a_l$ with a two-body interaction. If $|\alpha\rangle$ denotes in general the hole states of $|\Psi\rangle$ and $|\bar{\alpha}\rangle$ the particle orbitals, it follows immediately from the Schrodinger equation that :

$$U(\Delta t)|\Psi\rangle = \left(1 + \frac{\Delta t}{i\hbar} H^{(0)}\right)|\Psi\rangle + \frac{\Delta t}{i\hbar}\sum_{\alpha_1}\sum_{\bar{\alpha}_1} H^{(1)}_{\bar{\alpha}_1 \alpha_1}\, a^+_{\bar{\alpha}_1} a_{\tilde{\alpha}_1} |\Psi\rangle$$
$$+ \frac{\Delta t}{i\hbar}\sum_{\alpha_1 \alpha_2}\sum_{\bar{\alpha}_1 \bar{\alpha}_2} H^{(2)}_{\bar{\alpha}_1 \bar{\alpha}_2 \alpha_1 \alpha_2}\, a^+_{\bar{\alpha}_1} a^+_{\bar{\alpha}_2} a_{\tilde{\alpha}_1} a_{\tilde{\alpha}_2} |\Psi\rangle \qquad (1)$$

where $U(\Delta t)$ is the evolution operator and $|\tilde{\alpha}\rangle$ the biorthogonal hole basis which satisfies $\langle \tilde{\alpha}_1 | \alpha_2 \rangle = \delta_{\alpha_1 \alpha_2}$ and that have been introduced to take into account a possible non-orthogonality of the occupied states in the considered Slater determinant $|\Psi\rangle$. In addition, the coefficients in the expansion (1) are given by :

$$H^{(0)} = \sum_{\alpha}\sum_{ij}\langle \tilde{\alpha}|i\rangle T_{ij}\langle j|\alpha\rangle + \frac{1}{2}\sum_{\alpha_1 \alpha_2}\sum_{ijkl}\langle \tilde{\alpha}_1|i\rangle \langle \tilde{\alpha}_2|j\rangle V_{ijkl}\langle k|\alpha_1\rangle\langle l|\alpha_2\rangle \qquad (2.a)$$

$$H^{(1)}_{\bar{\alpha}_1 \alpha_1} = \sum_{ij}\langle \bar{\alpha}_1|i\rangle T_{ij}\langle j|\alpha_1\rangle + \sum_{\alpha_2}\sum_{ijkl}\langle \bar{\alpha}_1|i\rangle \langle \tilde{\alpha}_2|j\rangle V_{ijkl}\langle k|\alpha_1\rangle\langle l|\alpha_2\rangle \qquad (2.b)$$

$$H^{(2)}_{\bar{\alpha}_1 \bar{\alpha}_2 \alpha_1 \alpha_2} = -\frac{1}{4}\sum_{ijkl}\langle \bar{\alpha}_1|i\rangle \langle \bar{\alpha}_2|j\rangle V_{ijkl}\langle k|\alpha_1\rangle\langle l|\alpha_2\rangle \qquad (2.c)$$

Furthermore, given any two-body interaction, it is always possible to find one body operators $Q_s$ so that the 2p-2h amplitude (2.c) is written as :

$$H^{(2)}_{\bar{\alpha}_1 \bar{\alpha}_2 \alpha_1 \alpha_2} = \frac{1}{4}\sum_s \hbar \omega_s \langle \bar{\alpha}_1 | Q_s | \alpha_1 \rangle \langle \bar{\alpha}_2 | Q_s^+ | \alpha_2 \rangle \qquad (3)$$

For example, a possible set $\{Q_s, \omega_s\}$ can be obtained from the eigenvalue equation $\sum_{k,l} V_{ilkj}\langle k|Q_s|l\rangle = \hbar\omega_s \langle i|Q_s|j\rangle$ associated to the diagonalization of the hermitian matrix $M_{(i,j)(k,l)} = V_{ilkj}$. Introducing $X_s = (Q_s + Q_s^+)/2$ and $P_s = (Q_s - Q_s^+)/2i$, the evolution (1) of the Slater state $|\Psi\rangle$ to first order in the time-step $\Delta t$ becomes :

$$U(\Delta t)|\Psi\rangle = Exp\left(\frac{\Delta t}{i\hbar}H^{(0)}\right) Exp\left(\frac{\Delta t}{i\hbar}\sum_{\alpha_1}\sum_{\overline{\alpha}_1} H^{(1)}_{\overline{\alpha}_1 \alpha_1} a^+_{\overline{\alpha}_1} a_{\tilde{\alpha}_1}\right)$$

$$\prod_s Exp\left[\frac{i\omega_s \Delta t}{4}\left(\sum_{\alpha_1}\sum_{\overline{\alpha}_1}\langle\overline{\alpha}_1|X_s|\alpha_1\rangle a^+_{\overline{\alpha}_1} a_{\tilde{\alpha}_1}\right)^2\right] Exp\left[\frac{i\omega_s \Delta t}{4}\left(\sum_{\alpha_1}\sum_{\overline{\alpha}_1}\langle\overline{\alpha}_1|P_s|\alpha_1\rangle a^+_{\overline{\alpha}_1} a_{\tilde{\alpha}_1}\right)^2\right]|\Psi\rangle$$

(4)

The dynamics can then be linearized with the help of a Hubbard-Stratonovitch transformation [17] allowing to interpret each evolution under the square of a 1p-1h operator as an infinite superposition of one-body evolutions, each of them in a fluctuating auxilliary field $\sigma$ distributed on a Gaussian weight. This transformation is in fact currently used in Monte-Carlo calculations of the equilibrium state of interacting-fermions systems at finite temperature [4]. However its direct application to real time problems leads to a complex distribution of the variables $\sigma$ [18] and we suggest to use as an alternative the following identity :

$$Exp(i\,x\,O^2) = \frac{1}{\sqrt{2\pi}}\int_{-\infty}^{+\infty} d\sigma\, Exp\left(-\frac{\sigma^2}{2}\right) Exp\left[(1 + i\,sign(x))\sqrt{|x|}\,\sigma\,O\right] \quad (5)$$

where $x$ is real and $O$ an operator. Defining the vector $\vec{\sigma}$ of all the fields $(\sigma_{X_s}, \sigma_{P_s})$ that are introduced by the path-integral (5) to linearize the evolution (4), and introducing the shorthand notation $d\vec{\sigma}\,G(\vec{\sigma}) = \prod_s \frac{d\sigma_{X_s} d\sigma_{P_s}}{2\pi} Exp\left(-\frac{\sigma_{X_s}^2 + \sigma_{P_s}^2}{2}\right)$, we finally obtain :

$$U(\Delta t)|\Psi\rangle = Exp\left(\frac{\Delta t}{i\hbar}H^{(0)}\right)\int d\vec{\sigma}\,G(\vec{\sigma})\,Exp\left(\sum_{\alpha_1}\sum_{\overline{\alpha}_1}\Delta Z_{\overline{\alpha}_1 \alpha_1}(\vec{\sigma})\,a^+_{\overline{\alpha}_1} a_{\tilde{\alpha}_1}\right)|\Psi\rangle \quad (6)$$

where

$$\Delta Z_{\overline{\alpha}_1 \alpha_1}(\vec{\sigma}) = \frac{\Delta t}{i\hbar}H^{(1)}_{\overline{\alpha}_1 \alpha_1} + \sum_s \sum_{O_s = X_s, P_s} \frac{1 + i\,sign(\omega_s)}{2}\sqrt{\Delta t\,|\omega_s|}\,\sigma_{O_s}\langle\overline{\alpha}_1|O_s|\alpha_1\rangle \quad (7)$$

But, according to Thouless's theorem [1], the evolution of the Slater determinant $|\Psi\rangle$ under any of the fluctuating 1p-1h Hamiltonians of the auxiliary field representation (6) gives a neighboring pure independent-particle vector whose hole states are linear combinations of the initial particle and hole orbitals. To first-order in $\Delta t$, the correlated wave function (6) can thus be brought, with the help of (2.a) and (2.b), in the form :

$$U(\Delta t)|\Psi\rangle = \int d\vec{\sigma}\, G(\vec{\sigma}) \prod_{\alpha} a^+_{\alpha+\Delta\alpha(\vec{\sigma})}|\ \rangle \tag{8}$$

with

$$|\Delta\alpha_1(\vec{\sigma})\rangle = \sum_{\alpha_2}|\alpha_2\rangle\, \Delta Y_{\alpha_2\alpha_1} + \sum_{\bar{\alpha}_1}|\bar{\alpha}_1\rangle\, \Delta Z_{\bar{\alpha}_1\alpha_1}(\vec{\sigma}), \tag{9}$$

the deterministic contribution $\Delta Y$ being defined by :

$$\Delta Y_{\alpha_1\alpha_2} = \frac{\Delta t}{i\hbar}\left[\sum_{ij}\langle\tilde{\alpha}_1|i\rangle\, T_{ij}\,\langle j|\alpha_2\rangle + \frac{1}{2}\sum_{\alpha_3}\sum_{ijkl}\langle\tilde{\alpha}_1|i\rangle\langle\tilde{\alpha}_3|j\rangle\, V_{ijkl}\,\langle k|\alpha_2\rangle\langle l|\alpha_3\rangle\right] \tag{10}$$

Together with the expression (7) of the $\vec{\sigma}$–dependent element $\Delta Z$, and after some algebraic manipulations, the variation (9) of the holes states can in fact also be written as :

$$|\Delta\alpha(\vec{\sigma})\rangle = \frac{1}{i\hbar}\left[T + \bar{V}(\rho) - \frac{1}{2}\rho\bar{V}(\rho)\right]|\alpha\rangle\,\Delta t + \sum_{s}\sum_{O_s=X_s,P_s}\frac{1+i\,sign(\omega_s)}{2}\sigma_{O_s}\sqrt{|\omega_s|\Delta t}\,(1-\rho)\,O_s\,|\alpha\rangle \tag{11}$$

where $\rho = \sum_{\alpha}|\alpha\rangle\langle\tilde{\alpha}|$ is the one-body density and $\bar{V}(\rho)$ the mean-field potential with matrix elements $\langle i|\bar{V}(\rho)|j\rangle = \sum_{kl}V_{ikjl}\,\rho_{lk}$. In the end, the full dynamics of an uncorrelated state $|\Psi\rangle$, during a small time-step $\Delta t$ under a one and two-body Hamiltonian, can be represented as the coherent average of Slater determinants that have evolved with a mean-field Hamiltonian supplemented with a fluctuating 1p-1h potential. This noise, which comes from the 2p-2h residual interaction, is in addition non-hermitian and depends linearly on external number-fields, each distributed on a unit gaussian weight. We also emphasize that the deterministic part differs from the standard Hartree-Fock approach by including the term $-\frac{1}{2}\rho\bar{V}(\rho)$ which in fact arises naturally to take into account the difference bewen the total mean-field energy and the sum of the HF eigenvalues [2]. Last but by no means least, the formulation (8)-(11) of the dynamics can be extended to many time-steps leading to a wave function at any time $t = N \Delta t$ that is always the expectation of random Slater determinants. The Fermi sea associated to these states is then evaluated by a recursive application of (11) with a different stochastic vector field $\vec{\sigma}^{\pm}$ at each iteration. Such an algorithm corresponds in fact to the discretized realization of a Markovian Langevin equation in the one-body Hilbert space in accordance to the strong Euler scheme [19]. In consequence, going to the continuous limit $\Delta t \to 0$, and denoting by $E(\mathtt{L})$ the expectation of a random functional, we can state that :

$$U(t) |\Psi\rangle = E\left( \prod_\alpha a^+_{\alpha(t)} \right) \qquad (12)$$

where the evolution of the hole states $|\alpha(t)\rangle$ is determined by the following Itô stochastic differential equation [19] with the initial condition $|\alpha(0)\rangle = |\alpha\rangle$:

$$d|\alpha(t)\rangle = \frac{1}{i\hbar}\left[ T + \overline{V}(\rho(t)) - \frac{1}{2}\rho(t)\,\overline{V}(\rho(t)) \right] |\alpha(t)\rangle\, dt$$

$$+ \sum_s \sum_{O_s = X_s, P_s} \frac{1 + i\,\mathrm{sign}(\omega_s)}{2} \sqrt{|\omega_s|}\, (1 - \rho(t))\, O_s\, |\alpha(t)\rangle\, dW_{O_s}(t) \qquad (13)$$

where $W_{O_s}$ refer to independent real Wiener processes with vanishing enesemble averages and that obey to the Itô stochastic calculus [20] :

$$E\left( dW_{O_s}(t) \right) = 0 \qquad (14.\mathrm{a})$$

$$dW_{O_{s_1}}(t)\, dW_{O_{s_2}}(t) = dt\; \delta_{O_{s_1} O_{s_2}} \qquad (14.\mathrm{b})$$

It is also important to specify that such a formulation provides a reinterpretation of the exact evolution of the N-body density operator $D(t)$ in terms of a mean over dyadics in which the bra and the ket are different Slater determinants $\prod_\alpha a^+_{\alpha(t)} |\ \rangle$, $\prod_{\alpha'} a^+_{\alpha'(t)} |\ \rangle$ with $|\alpha(t)\rangle$ and $|\alpha'(t)\rangle$ evolving independently under the stochastic Hartree-Fock equation (12) :

$$D(t) = U(t) |\Psi\rangle \langle\Psi| U^+(t) = E\left( a^+_{\alpha_1(t)} ... a^+_{\alpha_N(t)} | \rangle \langle | a_{\alpha'_N(t)} ... a_{\alpha'_1(t)} \right) \quad (15)$$

Similarly, the expectation value of any observable $A$ will be expressed by :

$$\langle A\rangle(t) = Tr[D(t)\,A] = E\left( \langle | a_{\alpha'_N(t)} ... a_{\alpha'_1(t)} A\, a^+_{\alpha_1(t)} ... a^+_{\alpha_N(t)} | \rangle \right) \quad (16)$$

The necessary use of pairs of stochastic uncorrelated wave functions to reconstruct the exact full density operator consitutes in fact the originality of the proposed approach in comparison with previous schemes [15] where a quantum incoherent propagation of states was assumed. By a totally different way and for interacting fermions systems, we thus end up with the same representation of the N-body density matrix as those obtained for the dynamics of Bose-Einstein condensates [21].

Finally, the stochastic mean-field interpretation (12)-(14) of the evolution process can be extended to imaginary time to obtain the ground state of the system. The only modification in the previous demonstration is now the use of the standard Hubbard-Stratonovitch transformation [17] which leads to the following representation of the Boltzmann operator when it acts on a Slater determinant $|\Psi\rangle = \prod_\alpha a^+_\alpha | \rangle$ :

$$Exp(-\beta H) |\Psi\rangle = E\left( \prod_\alpha a^+_{\alpha(\beta)} \right) \quad (17)$$

with $|\alpha(0)\rangle = |\alpha\rangle$ and

$$d|\alpha(\beta)\rangle = \left[ T + \overline{V}(\rho(\beta)) - \frac{1}{2}\rho(\beta)\,\overline{V}(\rho(\beta)) \right] |\alpha(\beta)\rangle\, d\beta$$

$$+ \sum_s \sum_{O_s = X_s, P_s} \frac{(1+sign(\omega_s)) + i\,(1-sign(\omega_s))}{2} \sqrt{\frac{\hbar|\omega_s|}{2}} (1-\rho(\beta))\, O_s\, |\alpha(\beta)\rangle\, dW_{O_s}(\beta)$$

$$(18)$$

The correlated many-body ground state can then be obtained in the limit of large $\beta$ where $Exp(-\beta H)$ bahaves like a filter eliminating all the overlaps with excited eigenstates. The stochastic one-body interpretation (17)-(18) is in fact particularly interesting since it never suffers from the sign problem which occurs in the Monte-Carlo calculations of expectations values in the canonical ensemble.

### III- Statistical fluctuations and stability of the stochastic approach

We now examine the growth of statistical errors in the stochastic sheme (12)-(13) or (17)-(18). This can be achieved by calculating the average of the norm of the deviation between the exact evolution and a stochastic realization. For time-dependent problems, we are thus interesting in the indicator : $\chi(t) = E\left(\langle\phi(t)|\phi(t)\rangle\right)$ with $|\phi(t)\rangle = U(t)|\Psi\rangle - \prod_\alpha a^+_{\alpha(t)}|\ \rangle$ where the single-particle wave-functions $|\alpha(t)\rangle$ evolve according to the Itô equation (13). Using Eq. (12) and the unitarity of the evolution operator, one then immediatly checks that :

$$\chi(t) = E\left(\langle\Psi(t)|\Psi(t)\rangle\right) - 1 \text{ with } |\Psi(t)\rangle = \prod_\alpha a^+_{\alpha(t)}|\ \rangle \tag{19}$$

Furthermore $\langle\Psi(t)|\Psi(t)\rangle$ corresponds to the determinant of the overlap matrix $g_{\alpha_1\alpha_2}(t) = \langle\alpha_1(t)|\alpha_2(t)\rangle$ associated to the Fermi sea of $|\Psi(t)\rangle$ and which evolves according to :

$$\begin{aligned} dg_{\alpha_1\alpha_2}(t) &= \langle d\alpha_1(t)|\alpha_2(t)\rangle + \langle\alpha_1(t)|d\alpha_2(t)\rangle + \langle d\alpha_1(t)|d\alpha_2(t)\rangle \\ &= \frac{dt}{2}\sum_s\sum_{O_s=X_s,P_s}|\omega_s|\langle\alpha_1(t)|O_s(1-\rho(t))O_s|\alpha_2(t)\rangle \end{aligned} \tag{20}$$

where the stochastic Hartree-Fock equations (13) and Itô's differentiation rules [20] have been used. In consequence the variation of $\chi$ during an infinitesimal time $dt$ is given by :

$$d\chi(t) = E\left(\langle\Psi(t)|\Psi(t)\rangle\ Tr(dg(t)\ g^{-1}(t))\right) \tag{21}$$

leading finally to :

$$d\chi(t) = \frac{dt}{2}E\left(\langle\Psi(t)|\Psi(t)\rangle\sum_s\sum_{O_s=X_s,P_s}|\omega_s|\sigma^2_{\Psi(t)}(O_s)\right) \tag{22}$$

where $\sigma_{\Psi(t)}(O_s)$ refer to the quantal fluctuations of the one-body operators $O_s$ in the uncorrelated stochastic wave function $|\Psi(t)\rangle$. Such a result then implies that the proposed approach will never explode in a finite dimensional one-body space : in this case, a trivial upper bound for the variance $\sigma^2_{\Psi(t)}(O_s)$ is in fact given by the square of the largest eigenvalue $\lambda_{O_s}$ of the observable $O_s$ and thus $d\chi(t) \leq \frac{dt}{2} \chi(t) \sum_s \sum_{O_s=X_s,P_s} |\omega_s| \lambda^2_{O_s}$, which after time integration gives :

$$\chi(t) \leq Exp\left(\frac{t}{2} \sum_s \sum_{O_s=X_s,P_s} |\omega_s| \lambda^2_{O_s}\right) \quad (23)$$

### IV- Simulation examples

As a first a step towards the use of the previous stochastic formulation in practical problems, we investigate some numerical implementations in exactly soluble models. More precisely, we consider a system of $\Omega$ fermions distributed among a number $n$ of energy orbitals, each of which is $\Omega$-degenerate. Single particle states are thus labelled by two numbers $i = 0, \text{L}, n-1$ for the level and $\omega = 1, \text{L}, \Omega$ to take into account degeneracy within each orbital. In addition, the Hamiltonian of the system is expanded onto the bilinear operators $G_{ij} = \sum_{\omega=1}^{\Omega} a^+_{i\omega} a_{j\omega}$ which generates a $U(n)$ Lie algebra : usually, it includes the one-body contribution $\sum_{i=0}^{n-1} \varepsilon_i G_{ii}$ of the individual levels with energies $\varepsilon_i$ and a two-body interaction written in terms of products of the generators $G_{ij}$. This non trivial model is analytically soluble by group theoretical techniques in some cases and mimics in particular the shell-model picture of the atomic nucleus. For $n = 2$ levels, it corresponds to the standard Lipkin-Meshkov-Glick model [22] which is usually used to check the validity of approximate many-body techniques. The three-orbital case has also been studied in particular in the semi-classical limit where the dynamics can exhibit a chaotic behavior [23].

We first focus on the two-level model with the following Hamiltonian :

$$H = \varepsilon J_z + V J_x^2 \qquad (24)$$

where the quasi-spin operators $J_x = (G_{10} + G_{01})/2$, $J_y = (G_{10} - G_{01})/2i$ and $J_z = (G_{11} - G_{00})/2$ have been introduced. The system is initialized in an excited Hartree-Fock state that we choose as $\prod_{\omega=1}^{\Omega} \left( \frac{1}{\sqrt{2}} a_{0,\omega}^+ - \frac{1}{\sqrt{2}} a_{1,\omega}^+ \right) | \ \rangle$. This vector diagonalizes $J_x$ with the lowest eigenvalue $m = -\Omega/2$ in the symmetric representation of $SU(2)$ characterized by an angular momentum quantum number $j = \Omega/2$. We then solve the stochastic Hartree-Fock equations (13) by a weak scheme [19] and use Eq. (16) to compute expectation values of some observables. The results are presented in Fig. 1 for $\Omega = 20$ particles and a strength interaction $\eta = \Omega V / \varepsilon = 2$. After an average over $5.10^6$ stochastic realizations, the expectation of the quasi-spin as well as those of the two and three-body operators $J_z^2$, $J_z^3$ are at all time in perfect agreement with the exact solution of the Schrödinger equation. Furthermore, even if we have observed that the associated error bars increase quasi-exponentially with time, they are less than 5 % at the final time thus showing the stability of the stochastic mean-field approach in the model.

We have also considered the three-level extension by examining the dynamics of a coherent state :

$$\left| \Psi(z_1, z_2) \right\rangle = \left( 1 + |z_1|^2 + |z_2|^2 \right)^{-\Omega/2} Exp(z_1 G_{10} + z_2 G_{20}) | \Psi_o \rangle = \prod_{\omega=1}^{\Omega} \frac{a_{0,\omega}^+ + z_1 a_{1,\omega}^+ + z_2 a_{2,\omega}^+}{\sqrt{1 + |z_1|^2 + |z_2|^2}} | \ \rangle \qquad (25)$$

where $\left| \Psi_o \right\rangle = \prod_{\omega=1}^{\Omega} a_{0,\omega}^+ | \ \rangle$ corresponds to the Slater determinant with all particles in the ground

orbital. The Hamiltonian includes a two-body interaction having equal matrix elements for raising or lowering a pair of particles :

$$H = \sum_{i=0}^{2} \varepsilon_i \, G_{ii} - \frac{1}{2} V \sum_{\substack{i,j=0 \\ i \neq j}}^{2} G_{ij}^2 \qquad (26)$$

In addition we choose equidistant levels ($\varepsilon_0 = 0$, $\varepsilon_1 = \varepsilon$, $\varepsilon_2 = 2\varepsilon$) and set $\eta = \Omega V/\varepsilon = 2$. For $\Omega = 10$ fermions equally shared among the three levels ($z_1 = z_2 = 1$) at the initial time, the evolutions of the mean number of particles in each orbital are shown in Fig. 2. Once again, the stochastic mean-field scheme is in position to reproduce the exact dynamics but with more statistical fluctuations than in the previous analysis. This is in fact easily understood because of the larger number of Brownian processes in the stochastic Hartree-Fock equations (13).

For spectroscopic aspects, the implementation of the propagation in imaginary time (18) has been realized in the two models (24) and (26), as shown in Fig.3. In both cases, by an average over many simulations to reduce statistical fluctuations, the stochastic one-body scheme rapidly converges to the exact binding energy.

Finally, to study the dynamics of an initial correlated state, we have investigated a coupling between the stochastic evolutions in real and imaginary time. In the two-level Lipkin model with the Hamiltonian (24), the system is supposed to be in its ground state at time $t = 0$ and we suddenly change the sign of the two-body interaction. From a practical point of view, starting from the uncorrelated wave function $\prod_{\omega=1}^{\Omega} a_{1,\omega}^+ \mid \rangle$, the stochastic Hartree-Fock representation of the Boltzmann operator (17)-(18) is first applied up to a sufficiently large value of $\beta$ to obtain the correlated ground state. Each stochastic Slater determinant thus obtained is then propagated in real time with the new Hamiltonian according to Eq. (13). An average over a set of such one-body realizations allows after all to reconstruct the exact many-body solution. This is confirmed by the results shown in Fig. 4 even if the control of statistical errors is more difficult than in the previous simulations.

## V- Conclusion

We have presented a new theoretical framework to solve exactly the N-body Schrodinger equation for a fermionic system with binary interactions. In particular, we have shown that the full dynamics of an uncorrelated state can be formultated in terms of the coherent average of Slater determinants that have evolved with a stochastic Hartree-Fock equation. The noise has been derived from a path-integral representation of the evolution operator and comes from the linearization of the 2p-2h residual interaction by a modified Hubbard-Stratonovitch transformation. In addition, a similar scheme has been proposed for the imaginary time propagation thus leading to a stochastic one-body interpretation of the ground state. Finally, the coupling with the fluctuating mean-field approach in real time allows to reach the exact dynamics of a correlated wave function. Statistical errors have also been investigated to show that the proposed stochastic equations never admit a divergent realization in a finite dimensional one-body space. The numerical implementation in exactly soluble models has confirmed the formal results : in all the cases that we have studied, the stochastic Hartree-Fock method when averaged over many simulations give the correct result with a reasonable growth of the statistical spread. We hope to address a first application to a realistic problem in the near future.

## Acknowledgements

O. J. acknowledges financial support from "Société de secours des amis des sciences".

**Figure captions**

**Fig. 1** : Time evolution of the expectation of some observables in the two-level Lipkin model with the Hamiltonian (24) for $\Omega = 20$ fermions and $\eta = \Omega V/\varepsilon = 2$. The initial state is the Slater determinant that diagonalizes $J_x$ with the eigenvalue $m = -\Omega/2$ in the space of angular momentum $j = \Omega/2$. The average over $5.10^6$ stochastic one-body simulations is compared to the exact dynamics and the standard mean-field approach. Error bars in the stochastic scheme are not drawn because they would be roughly of the size of the points.

**Fig. 2** : Stochastic mean field dynamics of the three-level Lipkin model in comparaison to the exact and the Hartree-Fock evolution. The Hamiltonian takes the simple form (26) and the parameters are given by $\varepsilon_0 = 0$, $\varepsilon_1 = \varepsilon$, $\varepsilon_2 = 2\varepsilon$ and $\eta = \Omega V/\varepsilon = 2$. We have considered $5.10^7$ simulations for $\Omega = 10$ particles in the initial coherent state (25) with $z_1 = z_2 = 1$.

**Fig. 3** : Mean energy as function of the inverse temperature $\beta$ in the stochastic Hartree-Fock simulation (17)-(18). In the two-level model, we consider the average over $10^7$ realizations for a system of $\Omega = 20$ fermions with the Hamiltonian (24) and the parameters $\varepsilon = 1, V = -0.5$. The initial state is the Slater determinant that diagonalizes $J_x$ with the eigenvalue $m = -\Omega/2$ in the space of angular momentum $j = \Omega/2$. For the three-orbital case, the Hamiltonian is given by Eq. (26) with $\varepsilon_0 = 0$, $\varepsilon_1 = 1$, $\varepsilon_2 = 2$, $V = 0.2$. The results corresponds to an average over $10^6$ stochastic simultations for $\Omega = 10$ fermions initially in the ground orbital.

**Fig. 4** : Dynamics of an initial correlated wave function in the two-level Lipkin model for $\Omega = 20$ fermions with the Hamiltonian (24). For $t < 0$, we suppose $\varepsilon = 1, V = -0.1$ and reconstruct the ground state using the stochastic Hartree-Fock equations (18) until $\beta = 5$ and with the wave function $\prod_{\omega=1}^{\Omega} a^+_{1,\omega} | \rangle$ as a starting point. At time $t = 0^+$, the interaction becomes suddenly $V = 0.1$ and we then apply the one-body evolution scheme (13) on each stochastic Slater determinant obtained with the propagation in imaginary time. An average over $10^8$ of such simulations is compared to the exact dynamics.

**Figure 1**

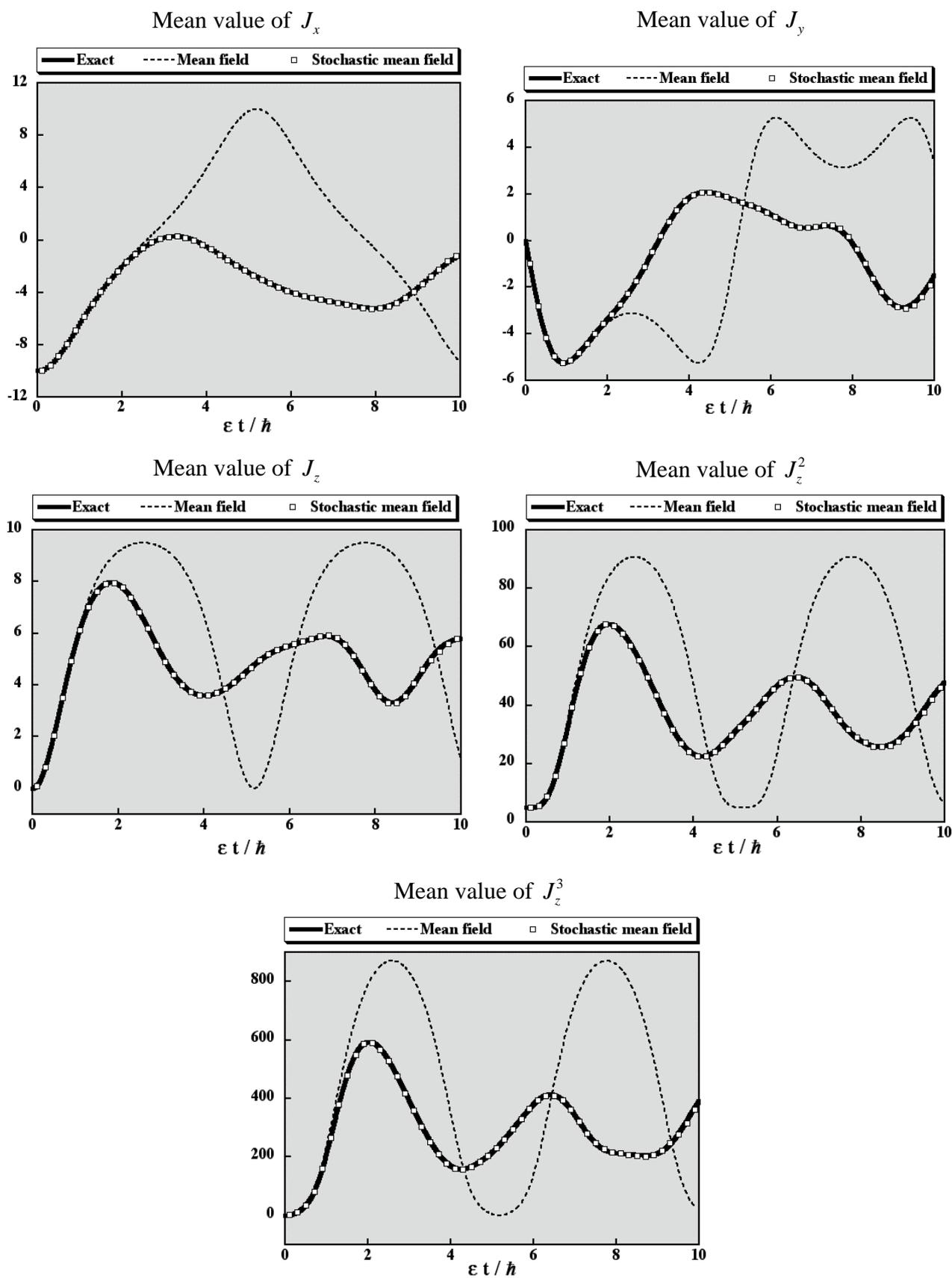

**Figure 2**

Mean particle number in the ground level

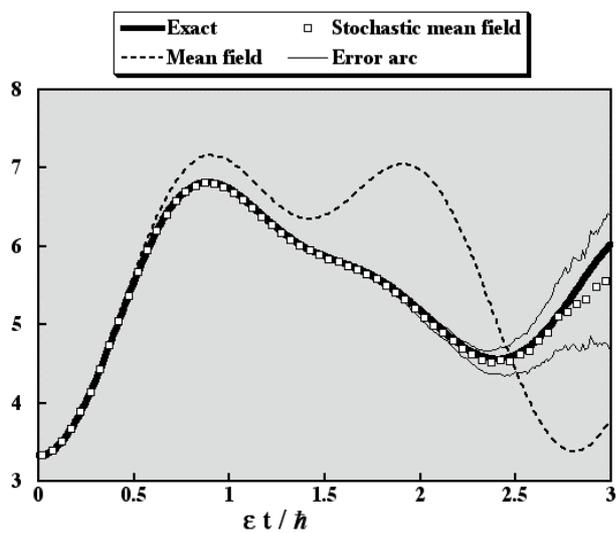

Mean particle number in the first excited orbital

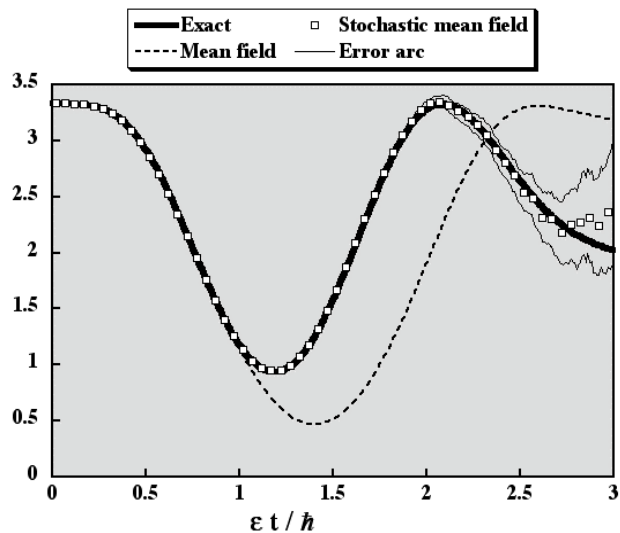

Mean particle number in the highest level

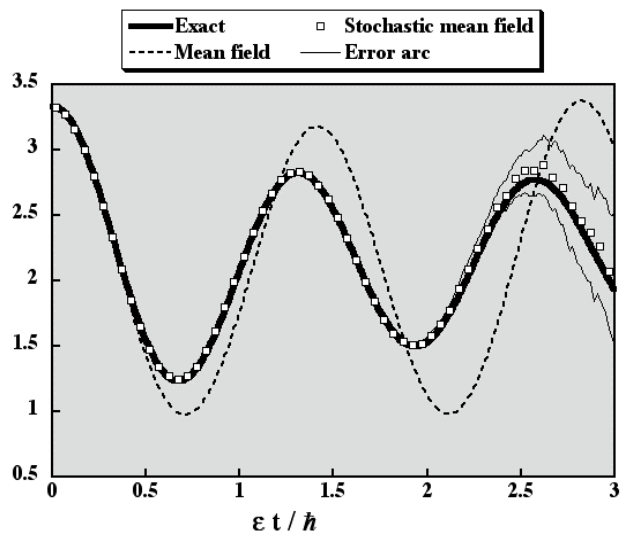

**Figure 3**

**Two-level Lipkin model**

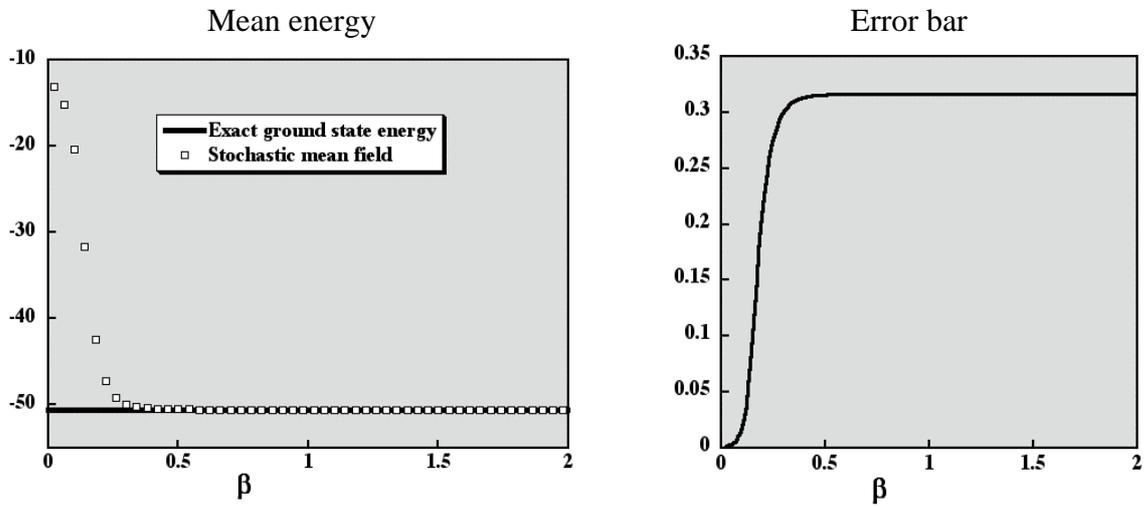

**Three-level Lipkin model**

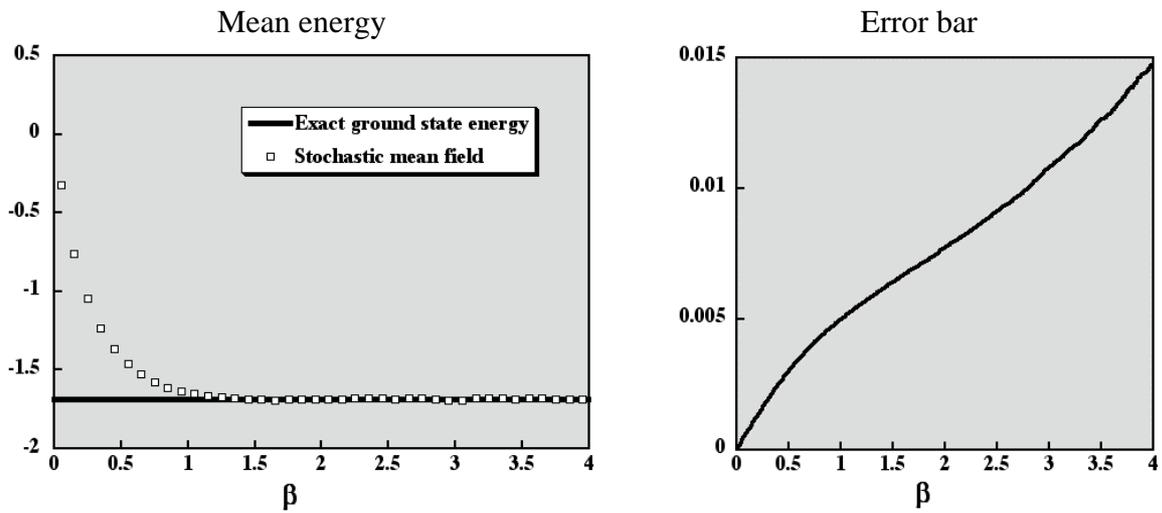

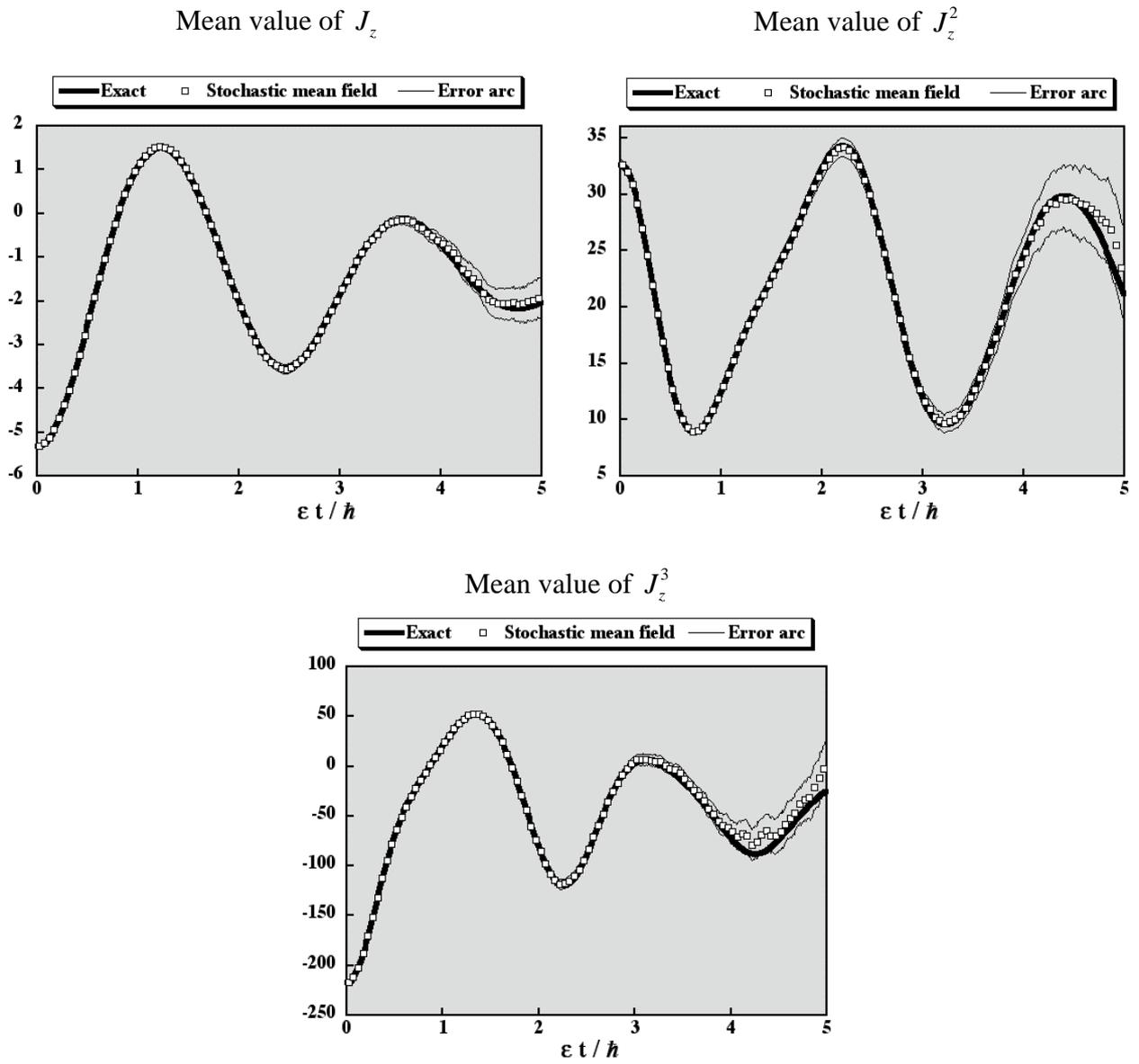